\documentclass[prb,aps,twocolumn,superscriptaddress,floats,showpacs,final, fleqn,hidelinks]{revtex4-2}
\usepackage{dcolumn}
\usepackage{bm}
\usepackage{color}
\usepackage{mathtools}
\usepackage{amsmath,amsfonts,amssymb,bm}
\usepackage{graphicx,tikz}
\usepackage[pdftex, bookmarks=true, linktoc=section, pdfstartview=FitH]{hyperref} 

\newcommand{\be}{\begin{equation}}
\newcommand{\ee}{\end{equation}}
\newcommand{\ba}{\begin{eqnarray}}
\newcommand{\ea}{\end{eqnarray}}

\begin{document}

\title{Transient Dynamical Phase Diagram of the Spin--Boson Model at Finite Temperature}

\author{Olga Goulko}
\email{olga.goulko@umb.edu}
\affiliation{Department of Physics, University of Massachusetts Boston, Boston Massachusetts 02125, USA}
\author{Hsing-Ta Chen}%
 \email{hchen25@nd.edu}
\affiliation{%
Department of Chemistry and Biochemistry, University of Notre Dame, Notre Dame, Indiana 46556, USA
}%

\author{Moshe Goldstein}
    \email{mgoldstein@tauex.tau.ac.il }
    \affiliation{
	School of Physics and Astronomy, Tel Aviv University, Tel Aviv 6997801, Israel
    }    

\author{Guy Cohen}
    \email{gcohen@tau.ac.il}
    \affiliation{
	School of Chemistry, Tel Aviv University, Tel Aviv 6997801, Israel
    }

\date{\today}

\begin{abstract}
We present numerically exact inchworm quantum Monte Carlo results for the real-time dynamics of the spin polarization in the sub-Ohmic spin--boson model at finite temperature. We focus in particular on the localization and coherence behavior of the model, extending our previous study at low temperature [Phys.~Rev.~Lett.~134, 056502 (2025)]. As the temperature increases, the system becomes less localized and less coherent. The loss of coherence, which is controlled by two independent mechanisms---a smooth damping-driven crossover and a sharp frequency-driven transition---exhibits a nontrivial temperature dependence. While both types of coherence loss occur at lower coupling in the high temperature regime, the frequency exhibits a sharper drop at high temperatures and this drop is observed for all values of the sub-Ohmic exponent, in contrast to the zero-temperature case. We discuss the full temperature-dependent dynamical phase diagram of the system and the interplay between coherence and localization across a wide range of physical parameters.
\end{abstract}

\maketitle

\section{Introduction}
One of the most foundational models in the study of open quantum systems is the spin--boson model (SBM)~\cite{leggettreview}.
The SBM comprises a single spin linearly coupled to a bosonic continuum with some spectral density.
This model and its extensions have been studied in various contexts going back to the 1960s. Some examples include its connection to the Kondo model, which is similar except in having a fermionic continuum \cite{guinea_diffusion_1985}; its applications to chemical dynamics in the presence of a dissipative environment \cite{nitzan_chemical_2006,quantumdissbook}, as well as electronic degrees of freedom coupled to optical cavities \cite{forn-diaz_ultrastrong_2019}; and as a paradigm for non-Markovianity in quantum information theory \cite{breuer_colloquium_2016}.

The SBM has numerous experimental implications and realizations, with some examples of recent interest coming on the one hand from using it to understand dissipation in certain types of qubits \cite{kaur_spin-boson_2021}, and on the other hand from using superconducting circuits \cite{drivenSBexp2017} or ultracold trapped ions \cite{sun_quantum_2025} to simulate it.

A particularly intriguing aspect of the SBM is the existence of a quantum phase transition (QPT) between the localized and delocalized phase when the low-energy behavior of the continuum spectral density obeys certain restrictions \cite{leggettreview}.
The spectral density can be parametrized by a continuous function $J(\omega)$ for $\omega>0$, often with power-law scaling at low energies, in the sense that $J(\omega) \propto \omega^s$ in some range of frequencies $0 \le \omega \lesssim \omega_\text{c}$ where $\omega_\text{c}$ is a cutoff frequency. At high energies, the spectral density often follows an exponential cut-off, $J(\omega) \propto e^{-\omega/\omega_c}$.
The Ohmic SBM ($s=1$) features a Berezinskii--Kosterlitz--Thouless transition between the localized and delocalized states, and the corresponding critical coupling, as well as the critical coupling for the onset of coherence (Toulouse point) are analytically known \cite{leggettreview, quantumdissbook}. The case $0<s<1$ is known as the sub-Ohmic regime, and will be the focus of the present study. This regime is of particular interest, on the one hand due to its rich phase diagram with a second order localization QPT and regimes of coherence and decoherence, and on the other hand due to its experimental relevance for superconducting and mesoscopic circuits \cite{tong2006mesoscopicring,yu2012supercondcircuits,drivenSBexp2017, leppakangas2018supercondircuit, yamamoto2019supercondcircuits} and trapped ion systems \cite{porras2008mesoscopic, lemmer2018trapped}. 

The critical behavior and phase diagram of the zero-temperature sub-Ohmic SBM in equilibrium are known. 
The infinite bandwidth limit can be understood from analytical considerations \cite{leggettreview}. 
More generally, our understanding rests on the quantum--classical correspondence between the SBM and an Ising spin chain with $1/r^{1+s}$ pair interaction \cite{anderson_exact_1970, yuval1970isingmapping, Anderson1971SomeNR, SBtoIsingCorresp}. 
Within this framework, the localization transition of the SBM is in the same universality class as the classical Ising transition and, therefore, the mapping predicts the structure of its equilibrium phase diagram and its critical scaling \cite{SBIsing1972,SBIsing1997}. 
The quantitative characterization of the equilibrium QPT has subsequently been refined by a variety of approaches. 
Early numerical renormalization group and renormalization group studies initiated a debate regarding the range of validity of the quantum--classical mapping and the associated critical exponents \cite{BullaEtAlCritExp2003,VojtaTongBulla2005,AndersBullaVojta2007}. 
Later work clarified the origin of the discrepancies \cite{VojtaTongBullaErratum2009, Vojta2012NRGerrors} and established consistency with the Ising-model predictions by employing improved renormalization-group treatments together with complementary equilibrium methods, including quantum Monte Carlo \cite{criticalExpSB}, exact diagonalization \cite{AlvermannFehske2009}, density matrix renormalization group approaches \cite{wong2008density}, variational matrix-product-state calculations \cite{GuoWeichselbaumDelftVojta2012}, and variational simulations with polaron ansatz \cite{entanglementSB2011,shen2023variational}. As a result, both equilibrium critical couplings and critical exponents have been determined reliably across the full sub-Ohmic regime.

The real-time dynamics in the sub-Ohmic regime has been studied with a variety of approximate and numerically exact methods. 
Early approximate approaches, including variational ans\"atze \cite{Chin_2006}, perturbative methods \cite{Lu_2007}, and flow equation calculations \cite{kehrein1996}, established a qualitative picture of coherent oscillations at weak dissipation and incoherent relaxation at stronger dissipation. 
These treatments also highlighted the renormalization of the tunneling element, its nontrivial temperature dependence, and the pronounced role of low-frequency bath modes in controlling decoherence. 
More detailed coherence maps and dynamical phase diagrams at zero temperature have been obtained using time-dependent variational wave-function approaches based on Davydov $\mathrm{D}_1$ trial states and their multi-$\mathrm{D}_1$ extensions \cite{wu_2013_davydov,yao_2013_davydov,wang2016subohmic,Lipeng2023}.

Beyond these approximate approaches, a number of numerically exact techniques have been applied to the dynamics of the SBM in general, including its sub-Ohmic regime \cite{Tanimura1989,Makri1992,makri_numerical_1995,wang2003,wang2008,wang2010coherent,Tanimura2005,YiJing2011,segal_nonequilibrium_2011,KastAnkerhold2013,Cao2013,Qiang2014,HEOM2017,subohmicCutoffDep2017,strathearn_efficient_2018,popovic_quantum_2021,cygorek_simulation_2022,gribben_exact_2022,Makri2023,grimm_accurate_2024,lackman-mincoff_path-filtering_2024,otterpohl2022sb,PhysRevLett.134.056502}. They provide complementary perspectives and benchmarks on transient relaxation, long-time equilibration, and coherence properties of the model. 
The dynamical view corresponds more closely to the time-dependent measurements performed in some of the experiments \cite{drivenSBexp2017,sun_quantum_2025}.
Interestingly, these studies indicate that the equilibrium localization transition and the loss of coherent oscillations in the dynamics need not coincide. Coherent oscillations can persist even in regions of the parameter space that are localized in equilibrium, while incoherent decay at zero temperature occurs only above a critical sub-Ohmic exponent $s_c$.
There is some disagreement regarding the precise value of $s_c$, but it is generally found to be in the range $0.4-0.5$ \cite{omegacDep2013,KastAnkerhold2013,KastAnkerholdPRB,wang2016subohmic,HEOM2017,otterpohl2022sb,Lipeng2023,PhysRevLett.134.056502}.

These transitions can be illustrated by sketching a transient dynamical phase diagram \cite{nalbach2010subohmic, PhysRevLett.134.056502}. 
Early numerical work by Nalbach and Thorwart \cite{nalbach2010subohmic} employed the quasiadiabatic propagator path integral (QUAPI) approach and showed that the low‑frequency bath modes generate a dynamical asymmetry, leading to a transient quasiequilibrium and ultraslow relaxation.
Note that these dynamical features of the sub-Ohmic SBM are unique to the initial condition where the state is factorized, but the bath is allowed to relax to an equilibrium state congruent with the initial state of the spin, which we will refer to as the \emph{shifted} bath case.
Another common choice is the \emph{unshifted} bath initial condition, where the state is factorized but the bath is at thermal equilibrium by itself. 
The dynamics of the sub-Ohmic SBM has been studied for both the shifted \cite{AndersBullaVojta2007,Lu_2007,takahashi2024heom,nalbach2010subohmic} and unshifted \cite{wang2008,wang2010coherent,HEOM2017,otterpohl2022sb,Sun_2016_variational,PhysRevLett.134.056502} initial conditions, including works that discuss both choices \cite{Tanimura2005,wu_2013_davydov,yao_2013_davydov,wang2016subohmic,KastAnkerhold2013,KastAnkerholdPRB,Lipeng2023}. Coherence properties of the model in general are highly sensitive to the choice of initial preparation.

Recently \cite{PhysRevLett.134.056502}, we mapped out the complete dynamical transient phase diagram of the QPT and the coherence regions of the sub-Ohmic SBM with the unshifted bath initial condition.
We used numerically exact short-time dynamics obtained from the inchworm Monte Carlo technique \cite{inchworm,thetainchworm1,thetainchworm2,antipov_currents_2017,dong_quantum_2017,ridley_numerically_2018,boag_inclusion-exclusion_2018,krivenko_dynamics_2019,ridley_numerically_2019,ridley_lead_2019,krivenko_dynamics_2019,chen_auxiliary_2019,eidelstein_multiorbital_2020,kleinhenz_dynamic_2020,erpenbeck_revealing_2021,yang_inclusion-exclusion_2021,li_interaction-expansion_2022,cai_fast_2022,pollock_reduced_2022,kleinhenz_kondo_2022,kim_pseudoparticle_2022,cai_inchworm_2023,erpenbeck_quantum_2023,erpenbeck_shaping_2023,cai_numerical_2023,wang_real-time_2023,cai_bold-thin-bold_2023,kim_vertex-based_2023,goldberger_dynamical_2024,kunzel_numerically_2024,atanasova_stark_2024,erpenbeck_steady-state_2024,strand_inchworm_2024,wang_solving_2025} and focused on the low-temperature limit.
The temperature dependence of the SBM dynamics in general has been studied in different contexts \cite{cao2000bathrelax,nalbach2010subohmic,lee2012sb,KastAnkerholdPRB,wang2017fintemp,wang2015sbmaster,Rosenbach2016sb,Sun_2016_variational,wu2017finitetemp,wang2017fcs,Yang2021sb,takahashi2024heom}.
For the shifted initial condition \cite{nalbach2010subohmic}, increasing temperature suppresses the dynamical asymmetry, and the incoherence crossover temperature increases with decreasing $s$. However, for the unshifted initial condition, the temperature dependence of the transient dynamical phase diagram has not yet been explored. Here we extend the dynamical phase diagram to the finite temperature regime.

The paper is organized as follows.
In Sec.~\ref{sec:model} we introduce the sub-Ohmic SBM and discuss our chosen parameters.
In Sec.~\ref{sec:method} we provide details regarding the computational scheme.
We then continue to the presentation and discussion of our data in Sec.~\ref{sec:results}, with additional data presented in Appendix \ref{sec:appendix}.
Finally, in Sec.~\ref{sec:conclusions} we conclude and summarize our findings.

\section{Model\label{sec:model}}
We consider the biasless SBM described by the Hamiltonian
\be
H=\frac{\Delta}{2}\hat{\sigma}_{x}+\frac{V_\mathrm{b}}{2}\hat{\sigma}_{z}+H_\mathrm{b},
\ee
where $\hat{\sigma}_x=|1\rangle\langle2|+|2\rangle\langle1|$ and $\hat{\sigma}_z=|1\rangle\langle1|-|2\rangle\langle2|$ are Pauli matrices and $\Delta$ is the tunneling amplitude.
The boson bath Hamiltonian is
\be
H_\mathrm{b}=\sum_{\ell}\omega_{\ell}(b_{\ell}^{\dagger}b_{\ell}+\frac{1}{2}),
\ee
where $b_\ell$ ($b^\dagger_\ell$) are the bosonic annihilation (creation) operators.
The system--bath coupling is assumed to be linear in the bath coordinates $x_\ell=\frac{1}{\sqrt{2\omega_{\ell}}}(b_{\ell}^{\dagger}+b_{\ell})$,
\begin{equation}
V_{\mathrm{b}} =\sum_{\ell}c_{\ell}x_{\ell}=\sum_{\ell}\frac{c_{\ell}}{\sqrt{2\omega_{\ell}}}\left(b_{\ell}^{\dagger}+b_{\ell}\right).
\label{eq:hybridization-hamiltonian}
\end{equation}
The coupling strength $c_\ell$ between the spin subsystem and the harmonic mode of frequency $\omega_\ell$ is characterized by the spectral density 
\begin{equation}
J\left(\omega\right)=\frac{\pi}{2}\sum_{\ell}\frac{c_{\ell}^{2}}{\omega_{\ell}}\delta\left(\omega-\omega_{\ell}\right)=2\pi\alpha\omega^s\omega_c^{1-s}e^{-\omega/\omega_c}.\label{eq:spectral}
\end{equation}
Here we choose the sub-Ohmic functional form where $\alpha$ controls the system--bath interactions, $\omega_c$ is the cutoff frequency, and $0<s<1$ is the sub-Ohmic exponent. As in Ref.~\onlinecite{PhysRevLett.134.056502} we use $\omega_c=10\Delta$ throughout.
For the initial condition, we consider the bath to be decoupled from the spin subsystem and the total density matrix takes the factorized form $\rho_0=|1\rangle\langle1|\otimes e^{-\beta H_b}/Z_b$ where $\beta=1/T$ and the bath configuration is chosen to be unshifted $Z_b=\textnormal{Tr}\left\{e^{-\beta H_b}\right\}$. 
Note that we set $\hbar=k_B=1$ throughout.

We are interested in the real-time dynamics of the spin subsystem, specifically in the time-dependent expectation value of the population difference between the two spin states,
\begin{equation}
    \langle \hat{\sigma}_z(t)\rangle =\textnormal{Tr}\left\{\rho_0e^{iHt}\hat{\sigma}_ze^{-iHt}\right\}.
    \label{eq:sigmazdyn}
\end{equation}
In Ref.~\onlinecite{PhysRevLett.134.056502}, we explored the behavior of the model near zero temperature ($T\rightarrow0$) with a focus on the localization and coherence behavior of the population as a function of coupling strength ($\alpha$) and sub-Ohmic exponent ($s$). 
In the present work, we turn our attention to extending this analysis to finite temperatures.

\section{Method and Parametrization\label{sec:method}}

To simulate the dynamics, we employed an inchworm quantum Monte Carlo (QMC) algorithm introduced in Refs.~\cite{thetainchworm1,thetainchworm2}.
Ref.~\cite{thetainchworm1} outlined two distinct expansions, and Ref.~\cite{thetainchworm2} applied them to an SBM with the Debye spectral density.
In Ref.~\cite{PhysRevLett.134.056502} we applied these expansions to the sub-Ohmic SBM in the $T\rightarrow0$ limit.
Here we provide a brief summary and details regarding the implementation in the finite-temperature sub-Ohmic case.

\subsection{System--bath coupling expansion}
The system--bath expansion scheme is built around perturbation theory in $V_\mathrm{b}$.
It is a continuous-time QMC algorithm \cite{gull_continuous-time_2011} analogous to the hybridization expansion method for fermionic impurity models.
While originally developed in imaginary-time for equilibrium problems \cite{werner_continuous-time_2006}, the hybridization expansion was rapidly applied to real-time dynamics \cite{muhlbacher_real-time_2008, schiro_real-time_2009, werner_diagrammatic_2009, schiro_real-time_2010, cohen_memory_2011}.
It was also the basis for one-shot bold methods \cite{gull_bold-line_2010,gull_numerically_2011,cohen_numerically_2013,cohen_greens_2014,cohen_greens_2014-1}, which reduced the dynamical sign problem limiting its application to longer times and lower temperatures; and eventually the first inchworm QMC scheme \cite{inchworm}, which completely overcomes the sign problem for most practical purposes.

The description of the bath in the method is fully parametrized by the autocorrelation function of the system--bath coupling,
\begin{equation}\label{eq:bath_correlation}
\begin{aligned}
    C(\tau)&=\langle V_\mathrm{b}(t+\tau)V_\mathrm{b}(t)\rangle_\mathrm{b}\\
    &=\frac{1}{2\pi}\int_0^\infty d\omega J(\omega)\left[\coth(\frac{\beta\omega}{2})\cos\omega\tau-i\sin\omega\tau\right]\\
    &\equiv Q_2(\tau)-iQ_1(\tau).
\end{aligned}
\end{equation}
Here $V_\mathrm{b}(t)$ denotes the time-dependence of the coupling in the interaction picture and $\langle\cdots \rangle_\mathrm{b}=\textnormal{Tr}\left\{e^{-\beta H_b}\cdots \right\}/Z_b$ denotes thermal averaging in the bath subspace at an inverse temperature of $\beta$.

\subsection{Diabatic coupling expansion}
The diabatic coupling expansion scheme, also introduced in Ref.~\cite{thetainchworm1}, is an expansion in the tunneling amplitude $\Delta$.
It is performed on a single time contour rather than a two-branch Keldysh contour, analogously to a Liouville equation.
In that sense, it is related to QMC schemes for fermionic systems where Keldysh indices have been summed over \cite{profumo_quantum_2015,bertrand_quantum_2019,macek_quantum_2020,bertrand_quantum_2021,nunez_fernandez_learning_2022,vanhoecke_diagrammatic_2023}.
To derive it, we first employ the Lang--Firsov polaron transformation to write the time evolution of the subsystem in terms of a product of two-time correlation functions.
We then reformulate the series as a cumulant expansion to enable taking advantage of the resummation properties inchworm algorithms depend on.
The key quantity in the parametrization of the method is then the autocorrelation function of the polaron shift operator,
\begin{eqnarray}
    \!\!\!\!\mathcal{C}(\tau)
    =\exp&\!&\left\{\frac{1}{2\pi}\int_0^\infty d\omega \frac{J(\omega)}{\omega^2}\coth(\frac{\beta\omega}{2})\right.\times\nonumber\\
    &\!&\ \ \ \left[(\cos\omega\tau-1)-i\sin\omega\tau\right]\biggr\}.
    \label{eq:cumulant}
\end{eqnarray}

\subsection{Bath correlation functions}
The two expansion schemes described above are somewhat complementary, in the sense that they exhibit distinct convergence properties within different parameter regimes \cite{thetainchworm2}.
We emphasize that the temperature dependence of the correlation function arises from the initial density matrix and can increase the maximal expansion order required to converge the dynamics.
That being said, the implementation of the inchworm algorithm is not affected by it in any conceptual way.

The bath correlation function is pre-computed using the following analytical tricks. For a spectral density with an exponential cutoff as given in Eq.~\eqref{eq:spectral}, the imaginary part of $C(\tau)$ can be evaluated analytically by
\begin{equation}
    Q_1(\tau)=\frac{\alpha\omega_c^2\Gamma(1+s)}{\left(1+\omega_c^2\tau^2\right)^{\frac{s+1}{2}}}\sin\left[(1+s)\tan^{-1}(\omega_c\tau)\right],
\end{equation}
where $\Gamma(z)=\int_0^\infty t^{z-1}e^{-t}dt$ denotes the gamma function. Note that the analytical expression reduces to $Q_1(\tau)=2\alpha\omega_c^3\tau/(1+\omega_c^2\tau^2)^2$ for $s=1$. 
However, for evaluating the real part $Q_2(\tau)$, there is a subtlety that arises from the sub-Ohmic spectral density. We notice that, in the limit $\omega\rightarrow0$, the integrand diverges as $\omega^s\coth(\frac{\beta\omega}{2})\rightarrow\frac{2}{\beta}\omega^{s-1}+\frac{\beta}{2}\omega^{s+1}+O(\omega^{s+3})$ if $0<s<1$. 
To avoid this problem, we split the integrand using $\coth(\frac{\beta\omega}{2})=\frac{2}{\beta\omega}+\left[\coth(\frac{\beta\omega}{2})-\frac{2}{\beta\omega}\right]$ and rewrite $Q_2(\tau)=Q_2^\prime(\tau)+Q_2^{\prime\prime}(\tau)$ where $Q_2^\prime(\tau)$ contains the divergent term $\frac{2}{\beta}\omega^{s-1}$. 
Fortunately, the Fourier transform of the divergent part of the integrand can be evaluated analytically by 
\begin{equation}
\begin{aligned}
    Q_2^\prime(\tau)=&\frac{2\alpha\omega_c^{1-s}}{\beta}\int_0^\infty \omega^{s-1} e^{-\omega/\omega_c}\cos(\omega\tau)d\omega  \\ 
    =& \frac{2\alpha\omega_c}{\beta}\frac{\Gamma(s)}{\left(1+\omega_c^2\tau^2\right)^{s/2}}
    \cos\left[s\tan^{-1}(\omega_c\tau)\right].
\end{aligned}
\end{equation}
We then numerically evaluate the remaining Fourier transform
\begin{equation}
\begin{aligned}Q_{2}^{\prime\prime}(\tau) & =\alpha\omega_{c}^{1-s}\int_{0}^{\infty}\omega^{s}e^{-\omega/\omega_{c}}\\
 & \times\left[\coth(\frac{\beta\omega}{2})-\frac{2}{\beta\omega}\right]\cos(\omega\tau)d\omega.
\end{aligned}
\end{equation}
This leads to a smooth form of $Q_2(t)$ which is suitable for numerical sampling and approaches the Ohmic form as $s\rightarrow1$.

The same trick is applied to the corresponding expression in the cumulant autocorrelation function in Eq.~\eqref{eq:cumulant}.
The integrand has an additional term of the form $1/\omega^2$ in this case; however, it is offset by the term $(\cos\omega\tau - 1)$ in the real part of the integrand, which scales as $\omega^2$ at small $\omega$.
Thus, the divergence has exactly the same form as in Eq.~\eqref{eq:bath_correlation}.

\begin{figure*}
    \centering
    \includegraphics[width=\textwidth]{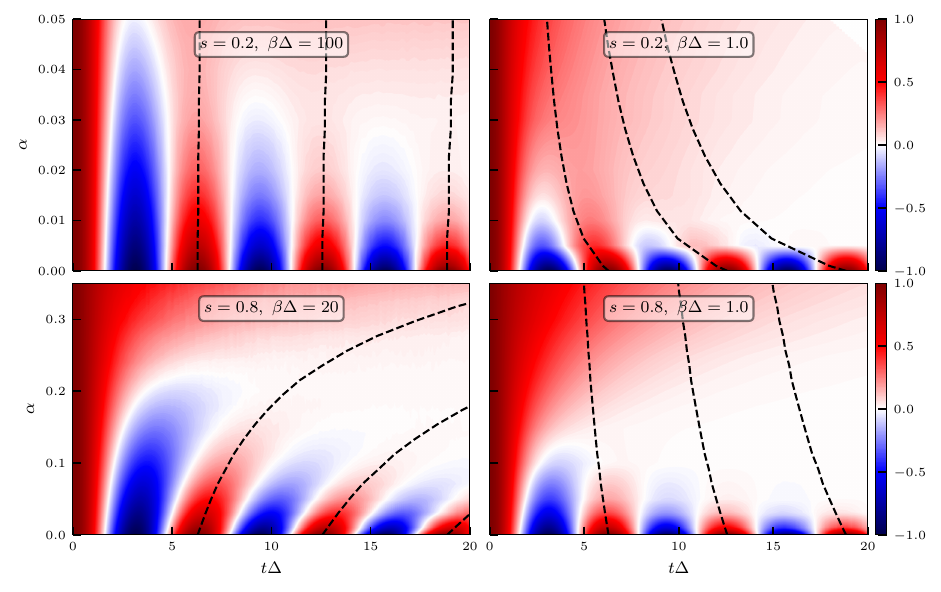}
    \caption{Time evolution of $\langle\sigma_z(t)\rangle$ as a function of coupling $\alpha$ deep in the sub-Ohmic regime (top panels) and close to the Ohmic regime (bottom panels) at low temperature (left panels) and at high temperature (right panels).
    Dashed black lines denote the analytical prediction for peaks in the dynamics, based on \cite{kehrein1996}.}
    \label{fig:tempdynamicsalpha}
\end{figure*}
\subsection{Fitting and Analysis}
Following Ref.~\onlinecite{PhysRevLett.134.056502}, we fit the dynamics of the subsystem $\langle\sigma_z(t)\rangle$ to the functional form
\begin{equation}
    \langle\sigma_z(t)\rangle \approx a\cos(\Omega t+\phi) e^{-\gamma_1 t} + be^{-\gamma_2t} + c\label{eq:fit}
\end{equation}
in order to extract several key characteristics. 
The damped oscillation at short timescales is described by the renormalized frequency $\Omega$ and the damping coefficient $\gamma_1$. 
The long-time behavior is captured by an asymptotic decay rate $\gamma_2$ (provided it is smaller than $\gamma_1$) and the offset coefficient $c$.

With the fitting coefficients extracted from Eq.~\eqref{eq:fit}, we can identify the following transitions and crossovers of interest:
(a) The transient dynamical delocalization transition occurs at $c=0$. We denote the corresponding critical coupling by $\alpha^*$, to distinguish it from the critical coupling $\alpha_c$ of the equilibrium quantum phase transition.
(b) The shift from coherence to incoherence is driven by two mechanisms: the renormalized frequency decreasing, which corresponds to the boundary of the $\Omega/\Delta=0$ region, and the over-damping of the oscillations, which we characterize as crossing the $\Omega=\gamma_1$ threshold. Since the latter is a smooth crossover, alternative definitions of the threshold are also possible.

It is worth noting that the localization parameter $c$ extracted from our heuristic fit function~\eqref{eq:fit} agrees well with recent independent estimates obtained using the purely data-driven ESPRIT (Estimation of Signal Parameters via Rotational Invariance Techniques) method~\cite{erpenbeck2025esprit}.

\section{Results\label{sec:results}}
\subsection{Time evolution of the spin polarization}
In Fig.~\ref{fig:tempdynamicsalpha}, we compare the dynamics of the subsystem at low temperatures ($\beta\Delta\gg1$, left panels) to the corresponding dynamics at high temperatures ($\beta\Delta=1$, right panels). Overall, the dynamics becomes more incoherent and more delocalized at higher temperatures.
Interestingly, we notice that there also appears to be a slight shift in frequency, which will be discussed in more detail below.
\begin{figure*}
    \centering
    \includegraphics[width=2\columnwidth]{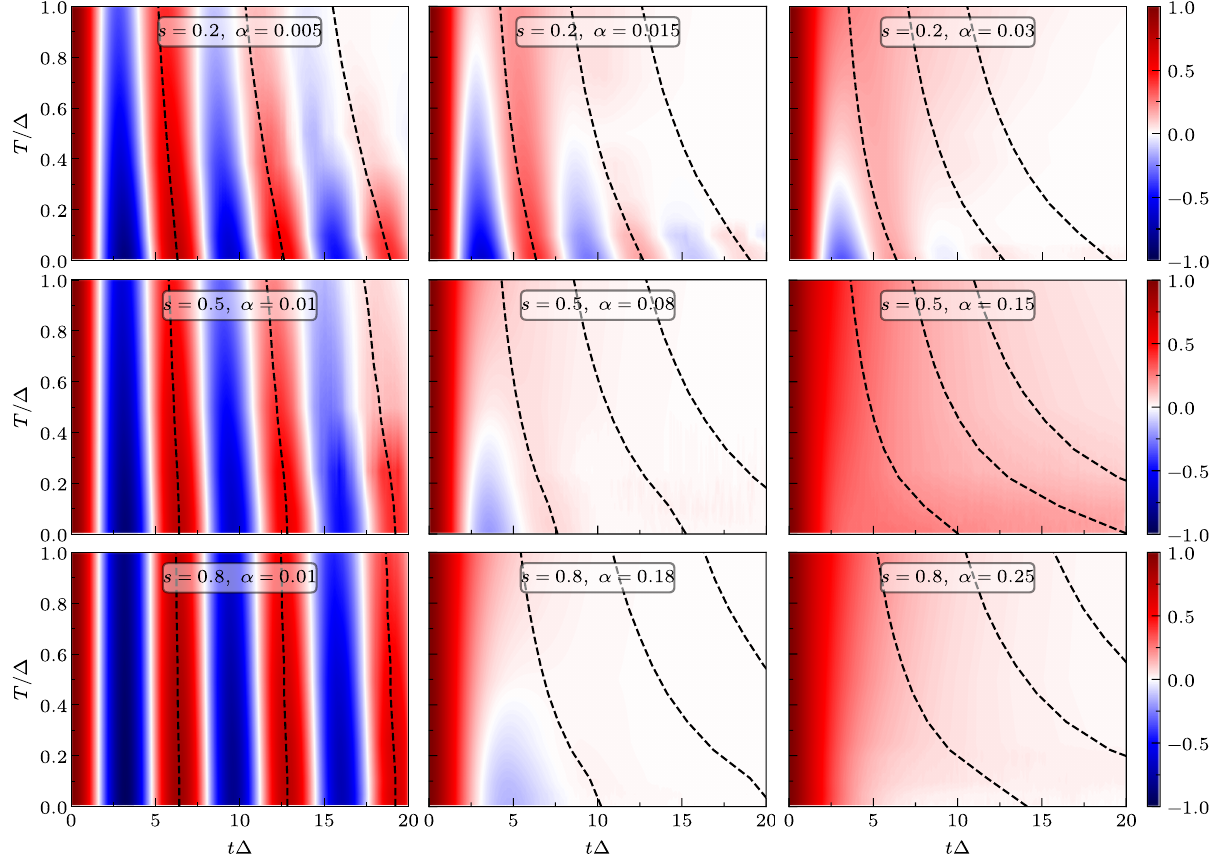}
    \caption{Time-evolution of $\langle\sigma_z(t)\rangle$ as a function of temperature $T$ deep in the sub-Ohmic regime at $s=0.2$ (top panels), at $s=0.5$ (center panels), and close to the Ohmic regime at $s=0.8$ (bottom panels) at different fixed values of coupling $\alpha$ (from left to right: below, near, and above $\alpha^*$).  Dashed black lines denote the analytical prediction for peaks in the dynamics, based on \cite{kehrein1996}.
    }
    \label{fig:tempdynamicsT}
\end{figure*}

It is illuminating to compare these numerical results to (almost) analytical predictions based on the flow equation theory of Ref.~\cite{kehrein1996}.
Here, as in Ref.~\cite{PhysRevLett.134.056502}, we solve the equation for the renormalized oscillation frequency numerically in terms of the exact bath spectral density $J(\omega)$, as given in Eq.~\eqref{eq:spectral}.
To visualize the analytical result, the dashed black lines in Fig.~\ref{fig:tempdynamicsalpha} represent multiples of the predicted frequency as a function of $\alpha$.

In the $\alpha=0$ limit (bottom edge of all panels), these curves exactly and trivially enumerate the maxima in the oscillating dynamics of the population.
At higher values of $\alpha$, they provide an approximate description.
Some of the trends are captured by the flow equations, especially in the weak coupling regime.
However, the discrepancy increases at higher temperatures and higher values of $\alpha$.

As discussed in Ref.~\cite{PhysRevLett.134.056502}, in the low-temperature regime, the subsystem dynamics transitions from a delocalized to a localized state and becomes less coherent with increasing coupling $\alpha$. Two independent decoherence mechanisms were identified: in the deep sub-Ohmic regime (cf.\ $s=0.2$, upper left panel), the loss of coherence is caused by the oscillation amplitude damping mechanism, without a significant change in frequency as $\alpha$ increases. Note that this damping mechanism is observed at all values of the sub-Ohmic exponent $s$.
However, at larger values of $s$ (cf.\ $s=0.8$, lower left panel), the loss of coherence is induced by an additional frequency-driven decoherence mechanism, i.e.\ a decrease in the oscillation frequency as $\alpha$ increases.

At higher temperatures (see the right panels of Fig.~\ref{fig:tempdynamicsalpha}) the damping-driven incoherence mechanism is much more strongly pronounced, as expected.
It is interesting to note that at high temperatures the flow equations predict an \emph{increase} in the frequency at stronger coupling.
This is supported by the numerical data in the deep sub-Ohmic weak-coupling regime (cf.\ $s=0.2$, upper right panel), but in general the numerically exact estimate for the frequency decreases with stronger coupling (cf.\ $s=0.8$, lower right panel). However, due to the pronounced damping at high temperatures, it is difficult to discern the oscillation frequency at higher values of $\alpha$ from the dynamics alone. A more detailed quantitative analysis based on the fit function Eq.~\eqref{eq:fit} will be provided in the following sections.
\begin{figure*}
    \centering
    \includegraphics[width=\textwidth]{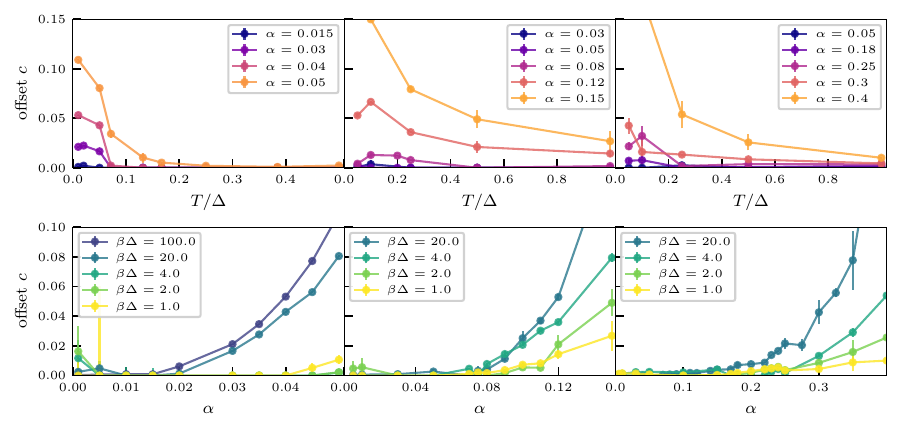}
    \caption{The offset fit coefficient $c$, which characterizes the localization transition. Top panels: Offset as function of temperature for different values of $\alpha$ for $s=0.2$ (left), $s=0.5$ (center), and $s=0.8$ (right). Bottom panels: Offset as function of $\alpha$ for different temperatures for $s=0.2$ (left), $s=0.5$ (center), and $s=0.8$ (right).}
    \label{fig:offset}
\end{figure*}

Figure~\ref{fig:tempdynamicsT} shows the dynamics of $\langle\sigma_z(t)\rangle$ once again, but this time for selected fixed values of $s$ and $\alpha$, and as a function of temperature $T=1/\beta$.
The top, middle and bottom rows show results with $s=0.2$, $s=0.5$ and $s=0.8$, respectively.
The left, middle and right columns show results below, near and above the zero-temperature dynamical delocalization transition $\alpha^*$, respectively.
Once again, the dashed black curves denote the flow equation prediction.

When the data are displayed in this fashion, it is immediately clear that raising the temperature generally increases the oscillation frequency (all panels).
This trend becomes less prominent at higher values of the coupling $\alpha$, where the strong damping obscures the oscillations.
The temperature-dependent change in frequency is generally well captured by the flow equations (dashed black curves), though the latter continue to predict the frequency even in regimes where damping makes it effectively unobservable.

As we noted in Sec.~\ref{sec:method}, we next fit these dynamics with a simple function Eq.~\eqref{eq:fit} in order to extract physically meaningful characteristics, including the long-time offset coefficient ($c$), the renormalized coherence frequency ($\Omega$), and the damping coefficient ($\gamma_1$).
This will enable us to present a wider view of how the dynamics depends on the various parameters, as well as a simpler picture of the physics.
The following subsections will therefore discuss the fitting parameters obtained in this way rather than the time dependence itself.
We show plots of additional fit parameters pertaining to the non-oscillating decay component of the evolution ($\gamma_2$ and $b$) in Appendix \ref{sec:appendix}.
For completeness, a data file with values of all fit parameters is openly available \cite{goulko2025data}.

\subsection{Localization properties}
Figure~\ref{fig:offset} displays the fitted offset coefficient $c$ (cf. Eq.~\eqref{eq:fit}) for different temperatures, couplings $\alpha$, and values of $s$.
The top row of panels shows the temperature dependence at constant coupling, while the bottom row shows the coupling dependence at constant temperature.
The left, middle and right columns correspond to $s=0.2$, $s=0.5$ and $s=0.8$, respectively.
We find that, with increasing temperature, the system tends to become less localized, as reflected by lower values of $c$.
On the other hand, with increasing coupling the system becomes more clearly localized.

It is interesting to note that while the absolute value of $c$ decreases with temperature, the critical coupling $\alpha^*$ for the transient dynamical localization transition---i.e.\ the minimal $\alpha$ value at which the offset is non-zero in the bottom row of Fig.~\ref{fig:offset}---is not characterized by such a clear trend.
With increasing temperature, the transition becomes less sharp (especially at higher values of $s$) and more difficult to accurately resolve from the numerical data.
That being said, the critical values $\alpha^*$ can be approximately extracted from plots like Fig.~\ref{fig:offset}.
The results of such an analysis will be discussed in more detail in Sec.~\ref{sec:phase_diagram}.

Readers may also note that, to some degree, it is possible to identify critical exponents in the data, though these are sufficiently clear only at lower temperatures.
The critical behavior obtained from dynamics at short times in this limit differs from its eventual equilibrium counterpart and was discussed in Ref.~\cite{PhysRevLett.134.056502}.
\begin{figure*}
    \centering
    \includegraphics[width=\textwidth]{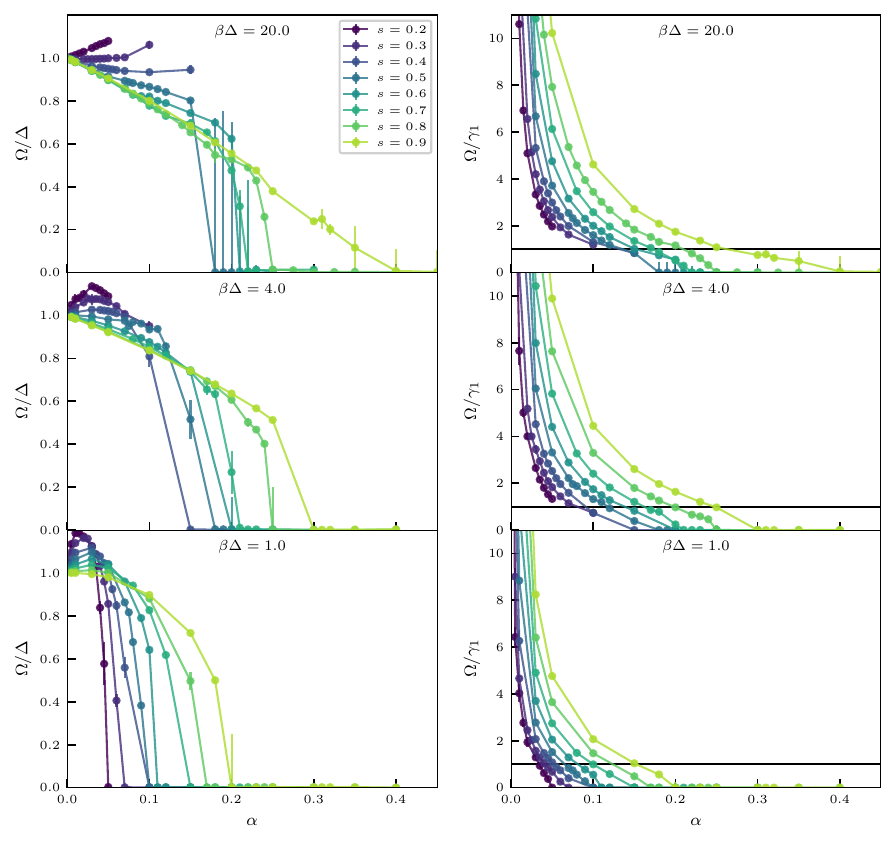}
    \caption{Left: the renormalized oscillation frequency, $\Omega/\Delta$. Right: the ratio between $\Omega$ and the corresponding damping coefficient $\gamma_1$. Temperatures (top to bottom): $\beta\Delta=20$, $\beta\Delta=4$, and $\beta\Delta=1$.}
    \label{fig:incoherence}
\end{figure*}

\subsection{Coherence properties}
Using our fit parameters, we next examine the temperature dependence of two hallmark signals indicating the loss of coherence.
The first of these is the decay of the frequency to zero, such that no evidence of oscillatory behavior is observed within the accessible timescale.
The second is the decrease in the ratio between the oscillation frequency and the damping rate.
In the latter case, even if some signature of oscillations at a finite frequency remains in the data, the dynamics becomes overdamped and, therefore, incoherent.
To further validate our results, we derive an analytical estimation of the damping rate in the low coupling limit.

\begin{figure*}
    \centering
    \includegraphics[width=\textwidth]{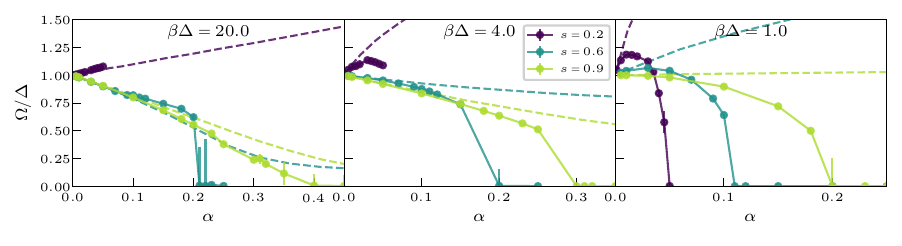}
    \caption{Oscillation frequency $\Omega$ as function of coupling $\alpha$ for three values of $s$ at different temperatures: $\beta\Delta = 20$ (left), $\beta\Delta=4$ (center) and $\beta\Delta = 1$ (right). The symbols represent numerical data (solid lines are guides to the eye) and the dashed lines of the corresponding color are the analytical predictions from \cite{kehrein1996}. The approximate theory and numerics agree at weak coupling, but the regime in which they agree becomes narrower at higher temperature. The numerically observed sharp drop to zero frequency at the frequency-drivel decoherence transition is not predicted by the theory.}
    \label{fig:freqwithRG}
\end{figure*}
\begin{figure*}
    \centering
    \includegraphics[width=\textwidth]{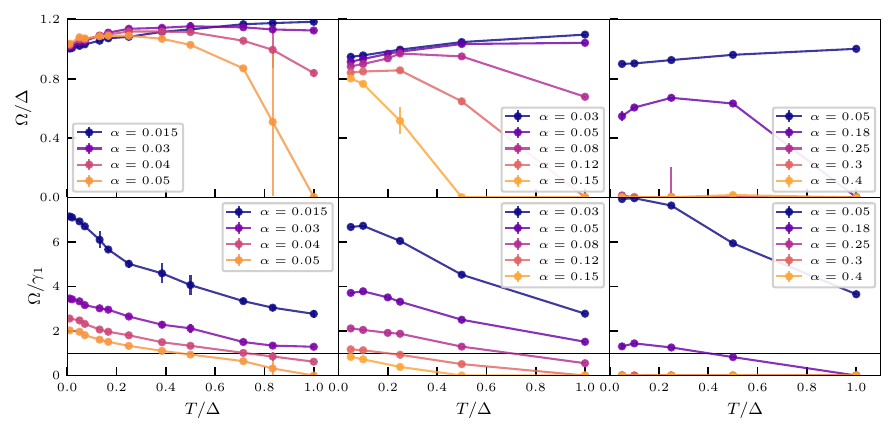}
    \caption{Top: the renormalized oscillation frequency, $\Omega/\Delta$. Bottom: the ratio between $\Omega$ and the corresponding damping coefficient $\gamma_1$. From left to right: $s=0.2$, $s=0.5$, $s=0.8$.}
    \label{fig:incoherence2}
\end{figure*}
\subsubsection{Coupling dependence}
In the left panels of Fig.~\ref{fig:incoherence} we present the oscillation frequency $\Omega$ in units of $\Delta$, and as a function of the coupling strength $\alpha$.
Each panel shows a series of exponents $s$ ranging from $0.2$, deep in the sub-Ohmic limit, to $0.9$, near the Ohmic regime.
From top to bottom the panels show data at increasing temperatures.
The frequency-driven loss of coherence is sharp and thus resembles the critical behavior at a phase transition, rather than a crossover.
This was discussed for low temperature in Ref.~\cite{PhysRevLett.134.056502}, but is even more apparent at higher temperatures, in contrast to the previously discussed critical behavior of the dynamical localization transition, which we found more difficult to observe at higher temperatures.
The only possible exception is at small values of $s$, where we cannot see frequency decay at accessible coupling strengths unless the temperature is very high.

Analogous plots of the ratio between oscillation frequency $\Omega$ and the damping coefficient $\gamma_1$ are presented in the right panels of Fig.~\ref{fig:incoherence}.
The thick horizontal lines denote an arbitrary criterion for the boundary between the coherent and incoherent regimes: $\Omega/\gamma_1=1$ (note that since $\gamma_1$ is always finite, demanding $\Omega/\gamma_1 \rightarrow 0$ would take us back to the first mechanism).
Displaying the data in this way provides a much simpler picture, where all the curves have a more similar functional form, and the analysis lends itself more easily to interpretation as a crossover from $\Omega/\gamma_1>1$ (coherence) to $\Omega/\gamma_1<1$ (incoherence).
Furthermore, the loss of coherence, here defined as overdamping, occurs for every value of $s$ and temperature that we examined.
\begin{figure*}
    \centering
    \includegraphics[width=\columnwidth]{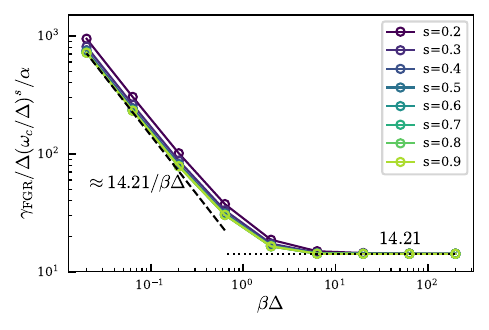}
    \includegraphics[width=\columnwidth]{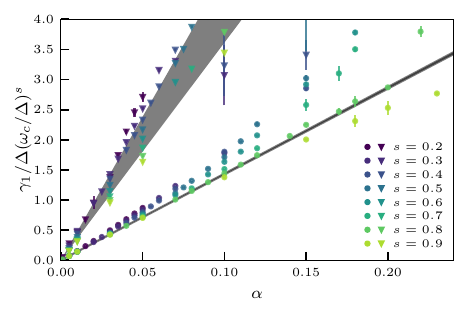}
    \caption{The damping coefficient of the oscillating dynamics for a range of sub-Ohmic exponents. Left panel: analytical prediction using Fermi's golden rule ($\gamma_\text{FGR}$) as a function of the inverse temperature $\beta$. The dashed and dotted lines denote the limits of small and large $\beta$, respectively. Right panel: numerical damping coefficient $\gamma_1$ for $\beta\Delta=20$ (circles) and $\beta\Delta=1$ (triangles). The shaded regions correspond to the analytical predictions, see the main text for discussion.}
    \label{fig:FGR}
\end{figure*}

As expected, coherence generally decreases when the temperature rises.
This applies to both decoherence mechanisms: the $\Omega/\Delta=0$ transition and the $\Omega/\gamma_1=1$ crossover both occur at lower values of $\alpha$ for all values of $s$.

The temperature dependence of the frequency-driven incoherence transition is particularly noteworthy.
In the left panels of Fig.~\ref{fig:incoherence}, we observe for some values of $s$ a sudden drop of the frequency to zero at higher values of $\alpha$.
Although at low temperatures the transition to $\Omega=0$ can only be observed in our data for $s\gtrsim0.5$, at higher temperatures all values of $s$ exhibit a sharp drop to $\Omega=0$ within the range of couplings accessible to our methodology.
This observation may explain the uncertainty in the literature about the critical sub-Ohmic exponent $s_c$ at low temperature \cite{omegacDep2013,KastAnkerhold2013,KastAnkerholdPRB,wang2016subohmic,HEOM2017,otterpohl2022sb,Lipeng2023}.
The comparison of the critical $\alpha$ values for the $\Omega/\Delta=0$ transition and the $\Omega/\gamma_1=1$ crossover will be discussed in more detail in Sec.~\ref{sec:phase_diagram}.

We also observe, in particular for smaller values of $s$, that in some cases an initial increase in $\Omega$ can be seen at lower values of $\alpha$.
At small values of $s$ and higher temperatures, this precedes the drop to $\Omega=0$.
The existence of the transition for $s \lesssim  0.5$ appears in conjunction with this nonmonotonic dependence of $\Omega$ with respect to $\alpha$.



This non-monotonic dependence of the frequency on $\alpha$ at weak coupling deep in the sub-Ohmic limit is also captured by the renormalization group calculations \cite{kehrein1996}, as Fig.~\ref{fig:freqwithRG} demonstrates.
Here, some of the data from the left panels of Fig.~\ref{fig:incoherence} is shown, along with the flow equation prediction (dashed curves).
Although the analytical predictions agree with the numerically exact results at small coupling $\alpha$, they start to deviate significantly at larger coupling and do not exhibit the sudden drop to zero frequency.
This discrepancy increases with increasing temperature.
Note that the renormalization group calculations are derived based on weak coupling and low temperature.  
This qualitative difference highlights that the numerically exact dynamics can capture the transition induced by increasing coupling, especially at finite temperatures. 

\subsubsection{Temperature dependence}
Figure~\ref{fig:incoherence2} shows the same observables presented in Fig.~\ref{fig:incoherence}, but as a function of temperature $T/\Delta$.
The upper row shows the frequency $\Omega/\Delta$, while the bottom row shows the ratio $\Omega/\gamma_1$ between frequency and damping.
The left, middle and right columns, respectively; show $s=0.2$, $s=0.5$ and $s=0.8$.
In each panel, several representative values of $\alpha$ are shown.

In this view, it is clear that increasing the temperature can either increase the frequency (at weak coupling and low temperatures) or decrease it.
At very weak coupling, no decrease in the frequency $\Omega/\Delta$ is observed at the range of temperatures we accessed; as a result, no phase-transition-like behavior is observed as a function of temperature.
When the coupling is strong enough, the frequency is already zero in the low temperature limit for $s\gtrsim0.5$.
Finally, at intermediate coupling strengths, the frequency goes to zero at some finite temperature in a critical-like manner.

On the other hand, $\Omega/\gamma_1$ always decreases with rising temperature, except at very low temperatures and weak coupling.
Here, the behavior looks very much like a crossover and no signature of criticality can be observed.

\begin{figure*}
    \includegraphics[width=\textwidth]{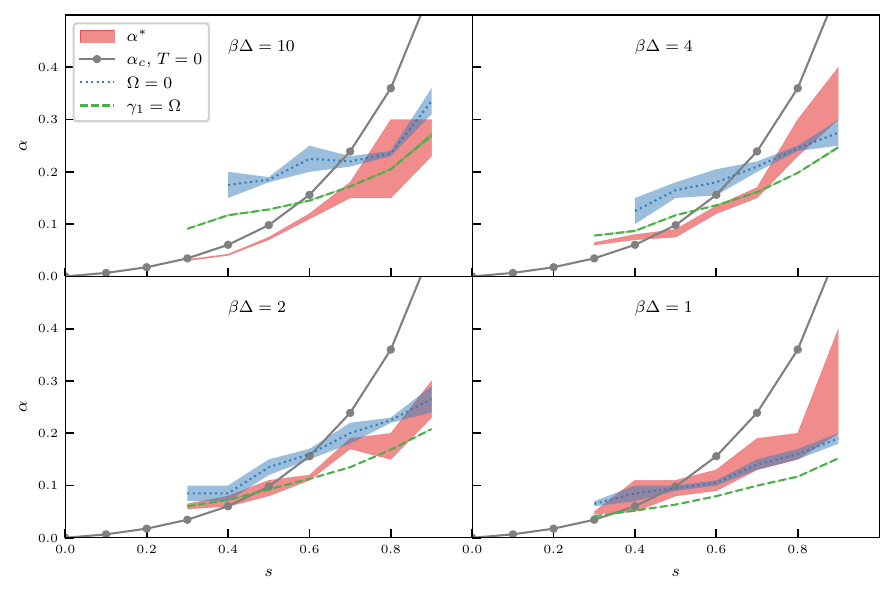}
    \caption{The phase diagram at successively increasing temperatures (see labels). Shown is the localization/delocalization transition $\alpha^*$, the $\Omega=0$ incoherence transition, and the $\Omega=\gamma_1$ incoherence crossover. For comparison, the zero temperature equilibrium $\alpha_c$ values from \cite{criticalExpSB} are included.\label{fig:phasediagseq}}
\end{figure*}
\subsubsection{Analytical expression of the damping rate}
In the low coupling limit ($\alpha\ll1$), one may
estimate the damping coefficient of the oscillating dynamics by Fermi's golden rule (FGR) \cite{nitzan_chemical_2006}: 
\ba
    \gamma_\mathrm{FGR}=\frac{1}{4}\int_{-\infty}^\infty dt e^{i\Delta t}C(t),
\ea
where $C(t)$ is the autocorrelation function of the system--bath coupling given by Eq.~\eqref{eq:bath_correlation}.
Since $\gamma_\mathrm{FGR}$ is proportional to $\alpha$, we can simply focus on the slope and carry out the time integration numerically. 
In the left panel of Fig.~\ref{fig:FGR}, we show that $\frac{\gamma_\mathrm{FGR}}{\Delta}\left(\frac{\omega_c}{\Delta}\right)^s/\alpha$ as a function of $\beta$ follow nearly identical behavior for all $s$: $\propto1/\beta$ for small $\beta$ and a constant for large $\beta$. 

We can approximate the asymptotic behaviors by the following analysis. 
For the low-temperature limit ($\beta\rightarrow\infty$ and $\coth(\frac{\beta\omega}{2})\approx1$), the damping rate can be approximated by \cite{PhysRevLett.134.056502}
\ba
\frac{\gamma_{FGR}}{\Delta}\left(\frac{\omega_c}{\Delta}\right)^s=\frac{\pi}{2}\alpha\frac{\omega_c}{\Delta}e^{-\Delta/\omega_c}\approx 14.21\alpha,
\ea
which agrees with the $\beta\rightarrow\infty$ limit in Fig.~\ref{fig:FGR}. 
For the high-temperature limit ($\beta\rightarrow0$ and $\coth(\frac{\beta\omega}{2})\approx\frac{2}{\beta\omega}$), we can approximate 
\ba\label{eq:gamma_FGR_highT}
\frac{\gamma_{FGR}}{\Delta}\left(\frac{\omega_c}{\Delta}\right)^s=\frac{\pi}{2}\frac{\alpha}{\beta\Delta}\frac{\omega_c}{\Delta}e^{-\Delta/\omega_c}\approx \frac{14.21}{\beta\Delta}\alpha,
\ea
which is depicted as the black dashed line in Fig.~\ref{fig:FGR}. 
We observe that the numerical integration results (labeled by open circles in Fig.~\ref{fig:FGR}) follow $1/\beta$ scaling for all $s$. 
Note that, while Eq.~\eqref{eq:gamma_FGR_highT} predicts a lower bound that does not depend on $s$, the numerical integration results are usually higher for small $s$ and agree with the approximate expression as $s$ increases.

To compare the predicted $\gamma_\mathrm{FGR}$ (as obtained by numerical integrations) and the damping coefficient $\gamma_1$ (as obtained by fitting the dynamics), we plot $\frac{\gamma_1}{\Delta}\left(\frac{\omega_c}{\Delta}\right)^s$ as a function of $\alpha$ for $\beta\Delta=20$ (circles) and $\beta\Delta=1$ (triangles) in the right panel of Fig.~\ref{fig:FGR}. 
The shaded areas correspond to the range of values assumed by the corresponding analytical expression, $\frac{\gamma_\mathrm{FGR}}{\Delta}\left(\frac{\omega_c}{\Delta}\right)^s$ between $s=0.2$ (upper) and $s=0.9$ (lower).
In the low temperature limit ($\beta\Delta=20$), since $\gamma_\mathrm{FGR}$ converges to the same value, the damping coefficient follows the predicted slopes $\approx14.21$.
For finite temperature ($\beta\Delta=1$), the predicted slopes range from $\approx36.13$ ($s=0.9$) to $\approx47.68$ ($s=0.2$) as shown in the left panel.
The spread of the numerical damping coefficients (triangles in the right panel) is consistent with these slopes at small $\alpha$.
At larger values of $\alpha$, the numerical data deviates from the theory as expected.

\subsection{Temperature dependence of the phase diagram}\label{sec:phase_diagram}
In the following, we discuss how the dynamical phase diagram obtained from the short-time relaxation of the sub-Ohmic SBM from the quench changes with increasing temperature.
We consider three different order parameters.
The first of these is the transient dynamical localization transition coupling $\alpha^*$, defined by the transition from zero to nonzero offset $c$ in Eq.~\eqref{eq:fit}. 
The second and third order parameters are the characteristics of the change from coherent oscillations to incoherent decay: the frequency transition line given by $\Omega=0$, and the damping-driven crossover defined as $\gamma_1=\Omega$.

Figure~\ref{fig:phasediagseq} shows the phase diagram at four different temperatures.
For reference, in each panel the solid gray line shows the equilibrium critical coupling $\alpha_c$ at zero temperature, as obtained in Ref.~\cite{criticalExpSB}.
In contrast to Ref.~\cite{PhysRevLett.134.056502}, where a similar phase diagram was shown at very low temperatures that were indistinguishable from the zero-temperature limit (specifically, $\beta\Delta=100$ for $0.2\leq s<0.5$ and $\beta\Delta=20$ for $0.5\leq s\leq1$),
the lowest temperature presented here is $\beta\Delta=10$.
We therefore do not expect more than a qualitative agreement between the equilibrium zero-temperature $\alpha_c$ and the value of $\alpha^*$ extracted from the dynamics, shown here in red with the span of the red area denoting our confidence intervals.
While the effect of temperature is small compared to the confidence interval in some of the parameter space, one trend is clear at least at small $s$: When the temperature increases, $\alpha^*$ also increases.
This might be expected, since temperature generally tends to delocalize the system.

\begin{figure}
    \includegraphics[width=\columnwidth]{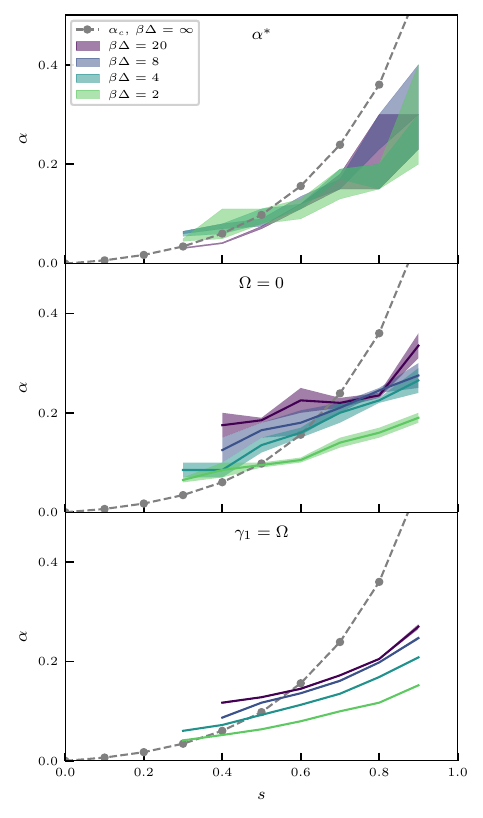}
    \caption{The phase diagram for localization transition (top), the frequency-driven incoherence transition characterized by $\Omega=0$ (center), and the damping driven incoherence crossover characterized by $\Omega=\gamma_1$ (bottom) for different temperatures. For comparison, the zero temperature equilibrium $\alpha_c$ values from \cite{criticalExpSB} are included.\label{fig:phasediagT_sequential}}
\end{figure}
\begin{figure}
    \includegraphics[width=\columnwidth]{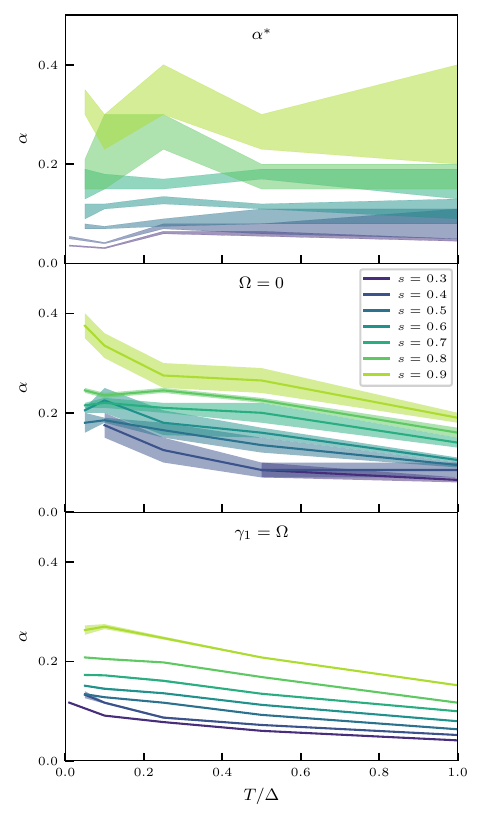}
    \caption{The phase diagram for localization transition (top), the frequency-driven incoherence transition characterized by $\Omega=0$ (center), and the damping driven incoherence crossover characterized by $\Omega=\gamma_1$ (bottom), with temperature as the horizontal axis.\label{fig:phasediagT}}
\end{figure}
\begin{figure*}
    \includegraphics[width=\textwidth]{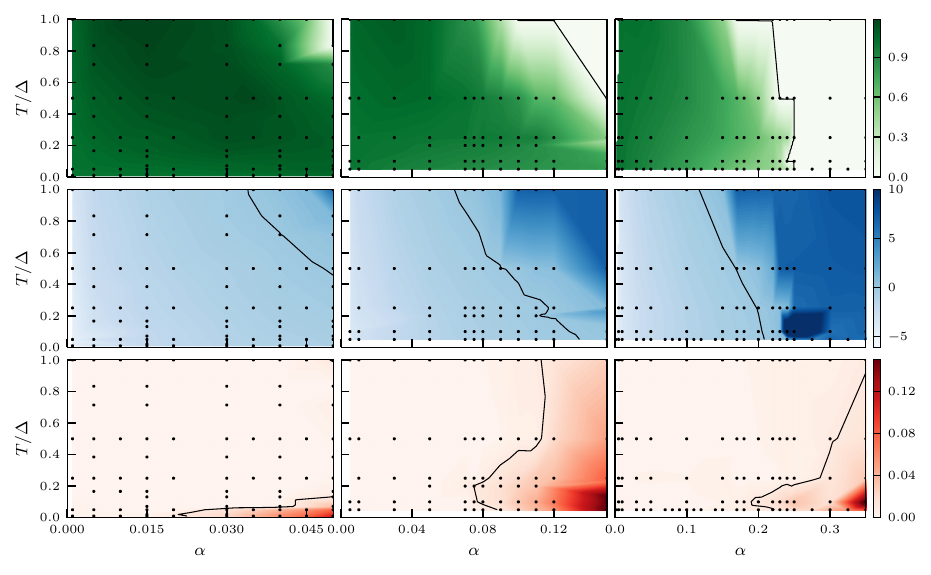}
    \caption{Contour plots of several dynamical fit parameters as a function of coupling ($\alpha$) and temperature ($T/\Delta$) for three representative values: $s=0.2,\ 0.5,\ 0.8$ from left to right. The color map indicates $\Omega/\Delta$ (top panels), $\ln(\gamma_1/\Omega)$ (center panels), and the offset $c$ (bottom panels). The dots indicate the parameter sets at which the inchworm QMC data were collected. The black contour lines correspond to the approximate locations of the respective transitions or crossovers: the frequency-driven incoherence transition line $\Omega=0$ (top panels), the damping-driven incoherence crossover line $\Omega=\gamma_1\Leftrightarrow\ln(\gamma_1/\Omega)=0$ (center panels), and the transient dynamical localization transition line $c=0$ (bottom panels).\label{fig:contourplots}}
\end{figure*}
The top panel of Fig.~\ref{fig:phasediagT_sequential} once again shows  $\alpha^*$, but with the results at several different temperatures overlaid.
This demonstrates more clearly that, other than the trend at small $s$, very little systematic dependence of this transition on temperature can be seen.

The two characteristics of the coherence--decoherence crossover are also shown in Fig.~\ref{fig:phasediagseq}.
The criterion $\Omega=0$ is denoted by a blue dotted curve, and has wide confidence intervals in some cases because it can be hard to distinguish exactly.
A green dashed curve similarly denotes $\gamma_1=\Omega$, which can be extracted much more accurately from our data.
The curves corresponding to both order parameters generally shift towards weak coupling as the temperature increases, such that the incoherent region is larger.
Once again, this is not particularly surprising because the temperature should act as a source of decoherence.

However, a nontrivial trend emerges in the relative locations of the respective transition and crossover lines: While at low temperatures the incoherence region is located almost entirely above the localization transition line in the localized phase, at high temperatures a more significant portion of the incoherence region (according to either incoherence criterion) falls below the localization transition line, or at least within the localization transition region.

The trends in the temperature dependence of the phase diagram for the coherence--decoherence crossover are more systematically observable from our data than those for the localization transition.
This is demonstrated more clearly in the middle and bottom panels of Fig.~\ref{fig:phasediagT_sequential} for $\Omega=0$ and $\gamma_1=\Omega$, respectively.
Once again, we point out that $\gamma_1=\Omega$ is substantially easier to evaluate at high accuracy, resulting in smaller confidence intervals and smoother data.

Although the dependence of the three order parameters we considered on temperature is relatively weak, their dependence on $s$ is strong and may make it difficult to see the effect of temperature.
In Fig.~\ref{fig:phasediagT} we therefore present the critical values for all three order parameters, in the same order from top to bottom as in Fig.~\ref{fig:phasediagT_sequential}; but here at constant $s$ and as a function of temperature $T$.
This reveals that the temperature dependence is somewhat stronger at large $s$, and displays more complicated behavior at low temperatures.
This suggests that investigating the low, but finite, temperature regime may be of interest in future studies if more precise numerical data can be obtained.

The critical (or crossover) values convey only some of the information contained within the fitting procedure for the dynamics.
In some cases, these values depend only weakly on the temperature, but the order parameters underlying them are more sensitive.
We therefore present a more comprehensive view of our order parameters as a set of contour plots, each plotted as a function of both the temperature and the coupling strength, in Fig.~\ref{fig:contourplots}.
The top, middle and bottom rows, respectively, show $\Omega/\Delta$, $\ln{(\gamma_1/\Omega)}$ and $c$.
The left, middle and right columns, respectively, show results at several representative exponents: $s=0.2$, $s=0.5$, and $s=0.8$.
Solid black contour lines appear at the threshold in each panel where we identify a transition or crossover.
These are meant mostly as a guide to the eye, and, therefore, to keep the plots less cluttered, no confidence intervals are shown for these lines here (the confidence intervals presented earlier were extracted from this data).

The top two rows show that the growth of the decoherence part of the phase space with temperature is clearly evident in both order parameters.
They also demonstrate that at high temperatures, the interesting physics occurs at rather high coupling strengths.
It would clearly be interesting, though computationally expensive, to map out the entire phase diagram in detail.
We leave this to future studies.

We have seen that the localization threshold does not exhibit a strong temperature dependence (see Figs.~\ref{fig:phasediagT_sequential} and \ref{fig:phasediagT}).
At the very least, any such dependence is difficult to identify given the large confidence intervals in our data.
It is therefore interesting to note that the temperature dependence of the offset (bottom panel of Fig.~\ref{fig:contourplots}) shows a significant and easily observable shift towards higher values at low temperatures and strong coupling.

\begin{figure*}
    \centering
    \includegraphics[width=\textwidth]{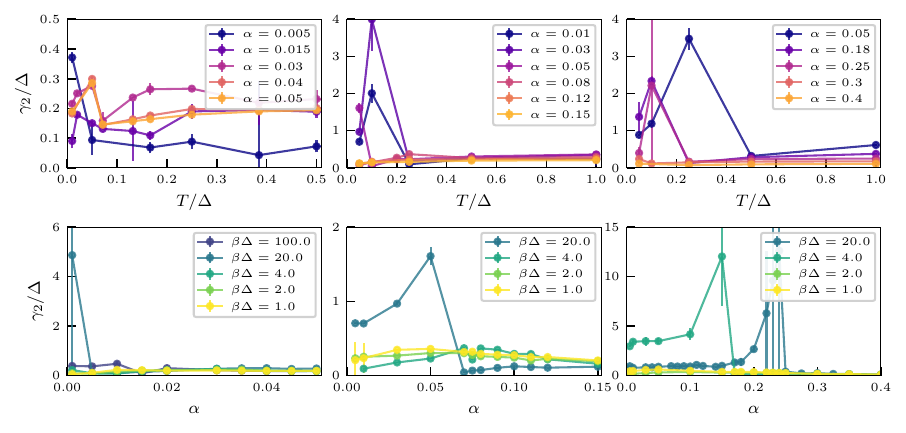}
    \caption{Fit coefficient $\gamma_2$, which characterizes the overall (non-oscillating) damping rate. Top panels: $\gamma_2/\Delta$ as function of temperature for different values of $\alpha$ for $s=0.2$ (left), $s=0.5$ (center), and $s=0.8$ (right). Bottom panels:  $\gamma_2/\Delta$ as function of $\alpha$ for different temperatures for $s=0.2$ (left), $s=0.5$ (center), and $s=0.8$ (right).}
    \label{fig:gamma2}
\end{figure*}
\begin{figure*}
    \centering
    \includegraphics[width=\textwidth]{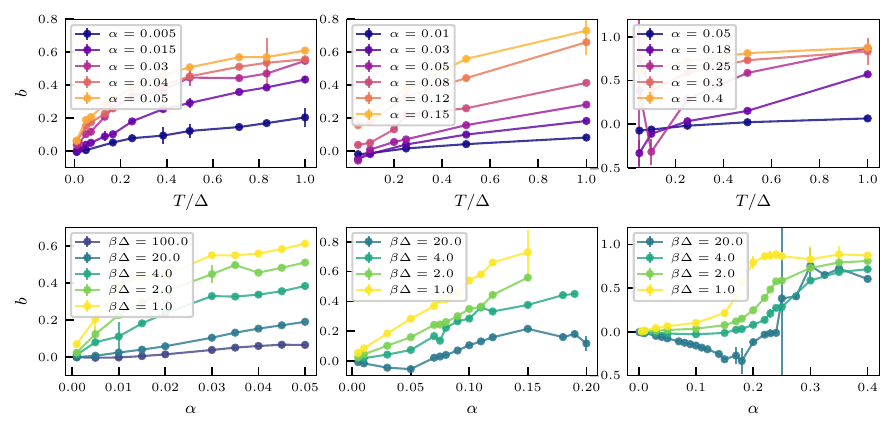}
    \caption{Fit coefficient $b$, which characterizes the amplitude of the overall (non-oscillating) damping term. Top panels: $b$ as function of temperature for different values of $\alpha$ for $s=0.2$ (left), $s=0.5$ (center), and $s=0.8$ (right). Bottom panels:  $b$ as function of $\alpha$ for different temperatures for $s=0.2$ (left), $s=0.5$ (center), and $s=0.8$ (right).}
    \label{fig:b}
\end{figure*}
\begin{figure*}
    \includegraphics[width=\textwidth]{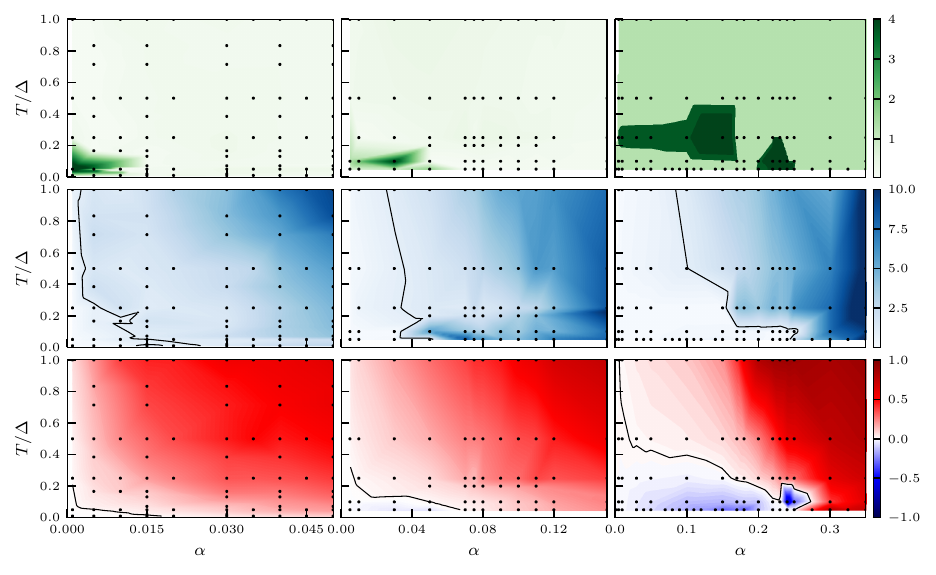}
    \caption{Contour plots of several additional dynamical fit parameters as a function of coupling ($\alpha$) and temperature ($T/\Delta$) for three representative values: $s=0.2,\ 0.5,\ 0.8$ from left to right. The color map indicates $\gamma_2/\Delta$ (top panels), the ratio $\gamma_1/\gamma_2$ (center panels), and the amplitude $b$ (bottom panels). The dots indicate the parameter sets at which the inchworm QMC data were collected. The black contour line in the center panels corresponds to $\gamma_1/\gamma_2=1$ and the black contour line in the bottom panels corresponds to $b=0$.\label{fig:contourplotsextra}}
\end{figure*}
\section{Conclusions\label{sec:conclusions}}
We have employed the inchworm QMC technique to obtain numerically exact short‑to‑intermediate–time dynamics of the sub‑Ohmic SBM at finite temperatures. Building on our earlier zero‑temperature study \cite{PhysRevLett.134.056502}, the present work systematically charts how temperature reshapes the transient dynamical phase diagram.

Finite temperature weakens both localization and coherent oscillations of the spin polarization.
The long‑time offset $c$ decreases essentially monotonically with temperature.
The critical coupling $\alpha^{*}$, where $c=0$, increases slightly with temperature for deep sub‑Ohmic exponents ($s \lesssim 0.4$), in accordance with the idea that thermal fluctuations should favor delocalization.
For larger $s$ the trend becomes less systematic, because the transition broadens and its numerical determination carries larger uncertainties.

Two different decoherence mechanisms were explored: (i) a smooth damping‑driven crossover ($\gamma_{1}=\Omega$) and (ii) a sharp frequency‑driven transition ($\Omega\!=\!0$).
Both mechanisms survive at $T>0$, but retreat towards weaker coupling as $T$ rises, so the incoherent sector of the phase diagram expands at the expense of the coherent one.
At $T/\Delta \gtrsim 0.5$, a substantial part of the incoherent regime already lies inside the delocalized phase, highlighting the fact that localization and coherence are distinct properties.

The predictions of the analytical flow equation reproduce the renormalized weak coupling frequency at all temperatures but do not capture the abrupt frequency collapse that signals the $\Omega=0$ transition.
The discrepancy becomes more prevalent as $T$ or $\alpha$ increases, underscoring the importance of numerically exact benchmarks.

At zero temperature and finite $\omega_c$ the sub-Ohmic SBM is known to exhibit a quantum phase transition between the localized and delocalized phases, which has been extensively studied in equilibrium or in the infinite time limit \cite{BullaEtAlCritExp2003,VojtaTongBulla2005, VojtaTongBullaErratum2009, Vojta2012NRGerrors,criticalExpSB,AlvermannFehske2009,GuoWeichselbaumDelftVojta2012,shen2023variational,HEOM2017}.
In our previous work \cite{PhysRevLett.134.056502}, we emphasized that the dynamical phase diagram extracted from the short- and intermediate-time behavior of the low-temperature model substantially differs from equilibrium physics.
However, experimental studies are performed at finite temperature.
Although the sharp boundary between phases is no longer present at finite temperature, one can nonetheless attempt to detect signatures of quantum critical behavior. 
This is often mentioned in the context of a ``quantum critical fan", which is an extended region in parameter space characterized by power-law temperature dependence of the observables \cite{Frerot2019cqf}.
For example, Ref.~\cite{AndersBullaVojta2007} schematically discusses the temperature evolution of the sub-Ohmic phase diagram and the emergence of a quantum critical region at finite temperature.

Although we do not see clear signatures of a quantum critical fan in this work, we observe that for the dynamical transient localization transition, the contour line for the offset $c$ as shown in Fig.~\ref{fig:contourplots} shows a trend that is directed opposite to the trend of the damping-driven incoherence contour line and, albeit to a lesser extent, the frequency-driven incoherence contour line.
The latter two contour lines are directed diagonally towards smaller values of the coupling with increasing temperature, while the localization contour tends to larger values of the coupling. 
The opposite temperature trends followed by the localization and decoherence contours hint at a narrow region where critical scaling could emerge.

In summary, we explored how temperature reshapes the dynamical phase diagram of the sub‑Ohmic SBM with unshifted bath initial conditions.
We found that it weakens localization, amplifies damping, and lowers coherence, maintaining the distinction between the two different decoherence mechanisms we considered. 
These findings could be potentially useful in guiding near‑term experiments that emulate spin--boson physics in superconducting circuits, trapped ions, or molecular nanomagnets.
More broadly, they embody an illuminating proof-of-concept for exploring the dynamical phase diagrams of quantum many-body systems and pose an interesting challenge to numerically exact techniques for simulating open quantum systems away from equilibrium.
The question of how---and perhaps in some cases whether---the transient phase diagram eventually decays to the equilibrium physics in the long-time limit remains open.

\section*{Acknowledgments}
We thank Jianshu Cao, Jan von Delft, Carlos Gonz\'alez-Guti\'errez, Olivier Parcollet, Nikolay Prokof'ev, and Andreas Weichselbaum for inspiring discussions.
O.G. is supported by the NSF under Grants No.~PHY-2441282, PHY-2112738 and OSI-2328774.
M.G. is supported by by the Israel Science
Foundation and the Directorate for Defense Research
and Development (DDR\&D) grant No. 3427/21, by the Israel Science Foundation grant No. 1113/23
and by the US-Israel Binational Science Foundation
(BSF) Grant No. 2020072.
G.C. is supported by the Israel Science Foundation (Grant No.~2902/21) and by the PAZY foundation (Grant No.~318/78).
We acknowledge high-performance computing support of the R2 compute cluster (DOI: 10.18122/B2S41H) provided by Boise State University's Research Computing Department, the chimera cluster provided by UMass Boston Research Computing, and the Unity cluster.
G.C. and HT. C. are supported by the TAU--ND Joint Research Program funded by the Schlindwein Family.

\appendix
\section{Fit parameters for the non-oscillating decay term}
\label{sec:appendix}
Figures~\ref{fig:gamma2}, \ref{fig:b}, and \ref{fig:contourplotsextra} show the temperature and coupling dependence of the fit parameters belonging to the second term in Eq.~\eqref{eq:fit}, which describes the non-oscillating part of the decay. Generally, the fit parameters in this term have larger error bars than the corresponding parameters in the oscillating term and the offset, because they are harder to precisely resolve numerically from short- and intermediate-time data. As can be seen in the center panels of Fig.~\ref{fig:contourplotsextra}, $\gamma_1>\gamma_2$ at larger couplings $\alpha$ and higher temperatures, hence the oscillating term is dominant at large times.

\bibliography{spinboson}

\end{document}